\documentstyle[preprint,tighten,eqsecnum,floats,epsfig,aps]{revtex}

\begin{document}

\draft

\title{ Chiral quark-soliton model in the Wigner-Seitz approximation }
\author{ P. Amore }
\address{
 College of William and Mary, Williamsburg, VA 23185, USA }
\author{ A. De Pace }
\address{
 Istituto Nazionale di Fisica Nucleare, Sezione di Torino, \\
 via P. Giuria 1, I-10125 Torino, Italy }
%\date{October 1999}

\maketitle

\begin{abstract}
In this paper we study the modification of the properties of the nucleon
in the nucleus within the quark-soliton model. 
This is a covariant, dynamical model, which provides a non--linear 
representation of the spontaneously broken $SU(2)_L\times SU(2)_R$ symmetry of 
QCD. The effects of the nuclear medium are accounted for by using the 
Wigner-Seitz approximation and therefore reducing the complex many-body problem
to a simpler single-particle problem.
We find a minimum in the binding energy at finite density, a change in
the isoscalar nucleon radius and a reduction of the in-medium pion decay 
constant. The latter is consistent with a partial restoration of chiral 
symmetry at finite density, which is predicted by other models. 
\end{abstract}

\section{ Introduction }

In this paper we want to address the possibility that the nucleon properties 
be modified in the nuclear medium. In a conventional nuclear physics 
approach, where nucleons and mesons are elementary degrees of freedom, such 
question cannot be answered consistently and any modification of the nucleon 
properties has to be put in by hand.
In order to address the problem, one has to consider models in which the 
substructure of the nucleon is not neglected (see, e.g, Refs.
\cite{Bah88,Rip97,Wal95}), and properly implement these models in order to 
account for the presence of a medium. 

In the present work we have considered a chiral model of the nucleon, which has
been developed by Diakonov {\em et al}. \cite{Dia88,Dia89} on the basis of the 
instanton picture of the QCD vacuum. It provides a low-energy approximation to
QCD that incorporates a non-linear representation of the spontaneously broken
chiral symmetry. 
In this framework pions emerge as Goldstone bosons, dynamically generated by 
the Dirac sea. Vacuum fluctuations (quark loops) are described by an effective 
action that yields the pion kinetic term, --- which is already included at the
classical level in the Lagrangian of other chiral models, --- and higher order 
non-local contributions. 
The model is also Lorentz covariant and has essentially only one free parameter
(apart from the regularization scale), namely the constituent quark mass.
Although the latter should in principle be momentum dependent, in practice a
constant value is usually chosen, which better reproduces the phenomenological
properties. The model has been successfully applied to the description of a
variety of nucleon properties \cite{Dia88,Dia89,Wak91,Dia98}.

Solving the many-body problem is already a formidable task in a conventional 
nuclear physics approach, the more so, of course, when one deals with extended
objects. Early attempts treated nuclear matter as a crystal, letting the 
particles sit on a regular lattice \cite{Nym70,Ach85,Kle85,Zha86}.
However, one has still to face serious computational difficulties in properly 
imposing the Bloch boundary conditions and moreover, nuclear matter does not 
show long-range crystalline order. These facts prompt the application of an 
approximation first introduced by Wigner and Seitz \cite{Wig33}, in which the 
effect of the surrounding matter on each particle is accounted in an average 
manner, by enclosing it in a spherically symmetric cell: This technique does 
not depend on any particular structure of the lattice and it is particularly 
suitable for nuclear matter, which may be pictured more as a fluid than a 
crystal.
In this way, a complex many-body problem is reduced to a single particle 
problem, where the effects of the nuclear medium enter only through average
boundary conditions.
The long-range order implied by imposing periodic boundary conditions gives
rise to a band structure of the energy levels and one has to choose suitable 
boundary conditions for the lowest and highest energy levels.

The Wigner-Seitz approximation to the treatment of soliton matter has already 
been applied using other models of the nucleon structure: The Skyrme 
model \cite{Wus87,Amo98}, non-topological soliton models 
\cite{Rei85,Bir88,Web98}, the hybrid soliton model
\cite{Ban85,Gle86,Hah87,Web98} and the global color model 
\cite{Joh96,Joh97,Joh98}.
In these models the boundary conditions for the spherically symmetric bottom
level of the band\footnote{ For the top level the problem is complicated by the
lack of spherical symmetry (see Ref.~\cite{Web98}). } are an extension to a
Dirac spinor of the non-relativistic requirement of having a flat wave
function. Moreover, in the chiral soliton models \cite{Web98,Ban85,Gle86,Hah87}
the requirement of unit topological number inside the cell is also taken into
account through the boundary condition on the chiral angle.

In this work we explore the sensitivity of the calculation to the choice of
different boundary conditions, by also imposing the requirement of flatness on
the chiral angle, as in Refs.~\cite{Nym70,Amo98}. 
We also show that it does not make sense to discuss the band structure of the 
nuclear system without accounting for the spurious contribution to the energy 
stemming from the center-of-mass motion of the bags, since the corrections turn
out to be larger than the band width and strongly dependent on the boundary 
condition for the level.

Another important difference of the present work from the above mentioned
calculations is the fact that here we also account for contributions coming
from the Dirac sea. Since the chiral field is a mean field, it means that we
include one-loop quark fluctuations. The latter generate the kinetic pion term,
--- which is included by hand in the model Lagrangians of
Refs.~\cite{Web98,Ban85,Gle86,Hah87}, --- and a (attractive) non-local
contribution, which is known to be important for the calculation of free
nucleon properties \cite{Dia88,Dia89,Wak91,Dia98}.
Dirac fluctuations are usually neglected in the in-medium calculations,
although there is no reason to expect they are negligible or at least
independent of the density.

The paper is organized as follows: In Sections~\ref{subsec:chimodel} and 
\ref{subsec:effact2} we briefly discuss the main features of the quark-soliton
model and the approximations employed in its implementation; 
in Sections~\ref{subsec:WS}--\ref{subsec:boundary} we introduce the 
Wigner-Seitz approximation, the appropriate boundary conditions for the fields
and show how a few observables can be calculated in this model. 
A new orthonormal and complete basis in the elementary cell is also obtained, 
in which physical quantities, such as the vacuum energy, are expressed. 
In Section~\ref{sec:res} we present the numerical results, obtained 
by solving the equations of motion.
Finally, in Section~\ref{sec:concl} we draw our conclusions and discuss
possible future developments of the model.

\section{ Nucleonic and nuclear models }
\label{sec:model}

\subsection{ Chiral quark-soliton model }
\label{subsec:chimodel}

The chiral soliton model \cite{Dia88,Dia89,Wak91}, which provides a non-linear
representation of the $SU(2)_L \times SU(2)_R$ symmetry of QCD, is based on 
the lagrangian
\begin{eqnarray}
  {\cal L} &=& \overline{\psi} \left[ i \rlap/\partial - M U_5(x) \right] \psi,
\label{eq:lagr}
\end{eqnarray}
where $\psi$ represents the quark fields, carrying color, flavor and Dirac 
indices, while $U_5$ is a chiral field defined as
\begin{eqnarray}
  U_5(x) &=& \frac{1 + \gamma_5}{2} \  U^\dagger(x) + 
    \frac{1 - \gamma_5}{2} \ U(x) \\
  U(x) &=& \exp\left[ i \bbox{\tau} \cdot \bbox{\theta}(x) \right] .
\label{eq:U5}
\end{eqnarray}
The large ($\cong350$ MeV) dynamical quark mass $M$ which appears in 
Eq.~(\ref{eq:lagr}) is the result of the spontaneous breakdown of chiral 
symmetry, which also accounts for the appearance of massless Nambu-Goldstone 
pions. In this model the nucleon emerges as a bound state of $N_c$ quarks in 
a color singlet state, kept together by the chiral mean field.
Note that no explicit kinetic energy term for the pion is present in
(\ref{eq:lagr}): Actually, the $\psi$ and $U$ fields are not independent and 
the latter is in the end interpreted as a composite field in the 
quark-antiquark channel.

Introducing the familiar hedgehog shape of the soliton,
\begin{eqnarray}
  U(x) &=& \exp\left\{ i \bbox{\tau} \cdot \hat{\bbox{r}} \ \theta(r)\right\} ,
\end{eqnarray}
the quark hamiltonian reads
\begin{eqnarray}
  H &=& - i \bbox{\alpha} \cdot \bbox{\nabla} + \beta M \left[ \cos \theta(r) 
    - i \gamma_5 \bbox{\tau}\cdot \hat{\bbox{r}} \sin \theta(r) \right] .
\label{eq:hamil}
\end{eqnarray}
However, solutions to the classical field equations derived from 
(\ref{eq:hamil}) do not
describe nucleon and $\Delta$ states, since angular momentum and isospin do not
commute with $H$: One defines the so called {\em grand spin}, 
$\bbox{K} = \bbox{J} + \bbox{\tau}/2$, for which $\left[H,\bbox{K}\right]=0$,
and the quark wave function can be written as
\begin{eqnarray}
  \psi &=& \frac{1}{\sqrt{4 \pi}} \left( 
    \begin{array}{c}
      u(r) \ \xi \\
      v(r) \ i \bbox{\sigma}\cdot \hat{\bbox{r}} \ \xi
    \end{array} \right) ,
\end{eqnarray}
where $\xi$ is the grand spin state fulfilling 
\begin{eqnarray}
  \left(\bbox{\sigma} + \bbox{\tau} \right) \ \xi = 0 
\end{eqnarray}
and the normalization is
\begin{eqnarray}
  4 \pi \int_0^\infty dr \ r^2 \ \overline{\psi} \gamma^0 \psi = 1 .
\end{eqnarray}

Good spin and isospin quantum numbers may be obtained in the end in a 
semiclassical approximation by quantizing the adiabatic rotational motion in 
isospin space \cite{Adk83,Rip97}.
However, in the present paper, for the sake of simplicity, we limit ourselves 
to consider soliton matter, leaving the projection of spin-isospin quantum 
numbers for future work.

The classical solutions are found self-consistently by solving the equations 
obtained by minimization of the total energy
\begin{eqnarray}
  \label{eq:Etot}
  E_{\text{tot}}[\psi,\overline{\psi},\theta] &=& 
    N_c E_{\text{val}}[\psi,\overline{\psi},\theta] + E_{\text{vac}}[\theta] ,
\end{eqnarray}
where $E_{\text{val}}\equiv\langle\psi|H|\psi\rangle$ and $E_{\text{vac}}$ are 
the valence and vacuum part of the energy, respectively. 
The vacuum energy $E_{\text{vac}}$ incorporates, at the mean field level, 
one-quark-loop contributions and, formally, can be evaluated through the 
effective action, which is obtained by considering the following path integral
over the quark fields 
\begin{eqnarray}
  \exp\left\{i S_{\text{eff}}[U] \right\} &=& \int [d\overline{\psi}][d\psi] \ 
    \exp\left\{i \int d^4x \ {\overline{\psi}(x) \left[ i \rlap/\partial - 
    M U_5 (x)  \right] \psi(x) } \right\} \nonumber \\
  &=& \left[ \det \left( i \rlap/\partial - M U_5(x) \right) \right]^{N_c} .
\label{eq:Seff1}
\end{eqnarray}
The latter can be easily cast in a more suitable form by means of simple 
algebraic manipulations 
\begin{eqnarray}
  S_{\text{eff}}[U] &=& -\frac{i}{2} N_c\, \text{tr}\, 
    \log\left[\frac{\Box + M^2+i M \rlap/\partial U_5(x)}{\Box+M^2} \right] ,
\label{eq:Seff2}
\end{eqnarray}
where the trace is over Dirac and flavor indices.
Despite its apparent simplicity, (\ref{eq:Seff2}) is actually a complicate 
nonlocal object. 

Although a local derivative expansion of course is possible, it is of little
practical use in this case, since the soliton field turns out to vary
significantly over the relevant distance scale $M^{-1}$ and no stable solutions
are found for expansions up to sixth order in derivatives \cite{Ait85}.
Kahana and Ripka \cite{Kah84} on the other hand have developed a numerical 
algorithm to directly evaluate vacuum polarization contributions to soliton
observables. This technique has been extended in Ref.~\cite{Wak91} to the
calculation of nucleon observables, that is after collective quantization has
been applied to project out states of definite spin and isospin.

\subsection{ Effective action up to second order in the soliton field }
\label{subsec:effact2}

Another path, to which we shall adhere in the following, has been followed in
Refs.~\cite{Dia88,Adj92} by expanding (\ref{eq:Seff2}) up to second order in
$M(\rlap/\partial U_5)$,  obtaining
\begin{eqnarray}
  S^{(2)}_{\text{eff}}[U] &\approx&  \frac{i}{4} N_c \, \text{tr} \,\langle x |
    \frac{1}{\Box + M^2} \left[i M \rlap/\partial U_5 \right] 
    \frac{1}{\Box + M^2} 
    \left[i M \rlap/\partial U_5 \right] | x \rangle \nonumber \\
  \label{eq:S2eff}
  &=& - \frac{1}{4} N_c \, \text{tr} \, \int d^4x d^4x' V(x) K(x,x') V(x') ,
\label{eq:Seffsecond}
\end{eqnarray}
where 
\begin{eqnarray}
  V(x)                 &=& i M \rlap/\partial U_5 \nonumber \\
  \label{eq:Kxx'}
  K(x,x')              &=& - i G(x,x') G(x',x)  \\
  (\Box + M^2) G(x,x') &=& \delta^4 (x-x') . \nonumber
\end{eqnarray}
Note that in the standard derivative expansion the second order action would 
read
\begin{eqnarray}
  S^{(2)}_{\text{eff}}[U] &\approx&  - \frac{i}{4} N_c M^2 \, \text{tr} \,
    \left[ \rlap/\partial U_5  \rlap/\partial U_5 \right] \langle x |
    \frac{1}{\Box + M^2} \frac{1}{\Box + M^2} | x \rangle .
\label{eq:Sefflocal}
\end{eqnarray}
In contrast to (\ref{eq:Sefflocal}) , Eq. (\ref{eq:Seffsecond}) does not assume
a slowly varying soliton field and gives rise to non-local contributions. 
Furthermore, one can see that it gives a good approximation both for small and 
large momenta, thus providing an interpolation formula between these regimes
\cite{Dia88,Dia89}.

Specializing to static field configurations, introducing 
$E^{(2)}_{\text{vac}}\equiv-S^{(2)}_{\text{eff}}/\int dx^0$ and going to
momentum space, one gets
\begin{equation}
  \label{eq:E2vac}
  E^{(2)}_{\text{vac}} = \frac{N_c}{4}\int\frac{d\bbox{q}}{(2\pi)^3}
    \text{tr}[V(\bbox{q})V(-\bbox{q})]K(q) ,
\end{equation}
where
\begin{equation}
  \text{tr}[V(\bbox{q})V(-\bbox{q})] = \frac{8M^2}{f_\pi^2}q^2
    [\phi_0(q)\phi_0(q) + \phi_i(\bbox{q})\phi_i(-\bbox{q})] ,
\end{equation}
with
\begin{mathletters}
\begin{eqnarray}
  \phi_0(q) &=& 4\pi f_\pi \int_0^\infty dr\,r^2 j_0(qr)[\cos\theta(r)
    -1] \\
  \phi_i(\bbox{q}) &=& i\hat{\bbox{q}}_i 4\pi f_\pi \int_0^\infty dr\,r^2 
    j_1(qr) \sin\theta(r) \equiv i\hat{\bbox{q}}_i \phi(q) ,
\end{eqnarray}
\end{mathletters}
and
\begin{eqnarray}
  K(q) &=& \int d\bbox{r}\,\text{e}^{i\bbox{q}\cdot\bbox{r}} K(r)
    \nonumber \\
  \label{eq:Kqr} \\
  K(r) &=& \frac{1}{8\pi^5} \int_0^\infty dk\,k^2 \int_0^\infty dk'\,{k'}^2
    \frac{\pi}{E_k E_{k'}} \frac{C\left[\left(k+k'\right)/2 
    \right]}{E_k+E_{k'}} j_0(kr) j_0(k'r) . \nonumber
\end{eqnarray}
In (\ref{eq:Kqr}) $E_k=\sqrt{k^2+M^2}$, whereas $C(k)$ is a regulating
function, which will be discussed later.

Although here and in the following we display, for convenience, formulae using
a momentum cut-off regularization scheme, we shall employ also the
Pauli-Villars regularization, also to be discussed later.

$E^{(2)}_{\text{vac}}$ must contain the {\em meson kinetic energy 
contribution}, which can be seen \cite{Adj92} to correspond to keeping 
only the $q=0$ term in an expansion of $K(q)$ in (\ref{eq:E2vac}). 
This requirement fixes the normalization of $K$ in such a way that 
\begin{equation}
  \label{eq:K(0)}
  K(0) \equiv \frac{1}{8\pi^2}\int_0^\infty dk\,k^2 \frac{C(k)}{E_k^3}
    = \frac{f_\pi^2}{4 N_c M^2} .
\end{equation}
Then one finds $E^{(2)}_{\text{vac}}=E^{\text{kin}}+\widetilde{E}^{(2)}$, where
\begin{eqnarray}
  E^{\text{kin}} &=& \frac{1}{4\pi^2}\int_0^\infty dq\,q^4
    [\phi_0^2(q)+\phi^2(q)] \nonumber \\
  &=& 2\pi f_\pi^2\int_0^\infty dr [r^2{\theta'}^2 + 2 \sin^2\theta] ,
\label{eq:Ekinfree}
\end{eqnarray}
with $\theta'=d\theta/dr$, and
\begin{equation}
  \widetilde{E}^{(2)} = \frac{1}{4\pi^2}\int_0^\infty dq\,q^4
    \left[\frac{K(q)}{K(0)}-1\right] [\phi_0^2(q)+\phi^2(q)] .
\label{eq:enl}
\end{equation}
The propagator $K(q)$ can be brought (see Appendix \ref{app:Kq}) into the form
\begin{equation}
  \label{eq:Kq}
  K(q) = \frac{1}{8\pi^2 q}\int_0^\infty dk\frac{k}{E_k}
    \int_{|k-q|}^{k+q} dk'\frac{k'}{E_{k'}} 
    \frac{C\left[\left(k+k'\right)/2\right]}{E_k+E_{k'}} .
\end{equation}

\subsection{ Nuclear matter in the Wigner-Seitz approximation }
\label{subsec:WS}

In order to describe nuclear matter we shall employ, as anticipated in the
Introduction, the Wigner-Seitz (WS) approximation \cite{Wig33}, which amounts
to enclose the fields in a spherically symmetric cell of radius R, imposing
suitable boundary conditions.
Before discussing our choice of boundary conditions, let us describe the
evaluation of the vacuum energy in the WS cell.

We have first to find an orthonormal and complete basis of functions inside the
elementary cell. We have chosen a spherical basis, in which the radial
dependence is expressed through spherical Bessel functions, which have
vanishing derivative at the boundary. This is the most useful basis in which to
perform the calculation, when {\em flat} (zero derivative) boundary conditions 
for the fields are invoked. More details on the basis are given in 
the Appendix \ref{app:basis}. All the quantities involved in the calculation of
the vacuum energy turn out to converge quickly using this basis.
They also converge rapidly when {\em zero} boundary conditions are employed
for the fields.

Starting again from the static limit of (\ref{eq:S2eff}) and (\ref{eq:Kxx'}),
one can introduce the Bessel transform of $K(\bbox{r},\bbox{r}')$ as 
\begin{equation}
  \label{eq:Krr'}
  K(\bbox{r},\bbox{r}') = \sum_{lm}\sum_{\alpha_{l}\alpha'_{l}}
    Y_{lm}(\hat{\bbox{r}}) Y^*_{lm}(\hat{\bbox{r}}') 
    \rho_{\alpha_{l}}(r) \rho_{\alpha'_{l}}(r') K_{l}(\alpha_{l},\alpha'_{l}) ,
\end{equation}
where
\begin{equation}
  \rho_{\alpha_{l}}(r) \equiv \kappa_{\alpha_{l}} j_{l}({\alpha_{l}}r/R) ,
    \quad \frac{d\rho_{\alpha_{l}}}{dr}\Big|_{r=R} = 0 ,
\end{equation}
$\kappa_{\alpha_{l}}$ being a normalization constant.
Inserting (\ref{eq:Krr'}) into (\ref{eq:S2eff}), one finds
\begin{equation}
  E^{(2)}_{\text{WS}} = E^{\text{kin}}_{\text{WS}} + 
    \widetilde{E}^{(2)}_{\text{WS}} ,
\end{equation}
where, as in the previous subsection, the kinetic energy contribution, --- the
one stemming from the local part of 
$K(\bbox{r},\bbox{r}')=\delta(\bbox{r}-\bbox{r}') K_0(r) + ...$, --- has been
separated. Indeed, one has
\begin{equation}
  E^{\text{kin}}_{\text{WS}} = 8\pi N_c M^2 \int_0^R dr K_0(r) 
    [r^2{\theta'}^2(r)+2\sin^2\theta(r)] ,
\end{equation}
with 
\begin{equation}
  \label{eq:Kr}
  K_0(r) = \sum_l\sum_{\alpha_{l}} \frac{2l+1}{16\pi} 
    \frac{C(\alpha_{l}/R)}{E^3_{\alpha_{l}/R}} \rho^2_{\alpha_{l}}(r)
\end{equation}
and $E_{\alpha_{l}/R}=\sqrt{(\alpha_{l}/R)^2+M^2}$.

In the limit $R\to\infty$ one can check that 
\begin{equation}
  K_0(r) \to \frac{1}{8\pi^2}\int_0^\infty dk\,k^2 \frac{C(k)}{E_k^3}
    \equiv \frac{f_\pi^2}{4 N_c M^2} .
\end{equation}
As in the free space case, this fixes the normalization of $K_0(r)$ in such a
way that 
\begin{equation}
  K_0(0) = \frac{1}{16\pi}\sum_{\alpha_{0}}
    \frac{C(\alpha_{0}/R)}{E^3_{\alpha_{0}/R}} \rho^2_{\alpha_{0}}(r) 
    = \frac{f_\pi^2}{4 N_c M^2} .
\end{equation}
Then, one can write
\begin{equation}
  \label{eq:EkinWS}
  E^{\text{kin}}_{\text{WS}} = 2\pi f_\pi^2 \int_0^R dr \frac{K_0(r)}{K_0(0)}
    [r^2{\theta'}^2(r)+2\sin^2\theta(r)] .
\end{equation}
As we shall discuss below, one can view
\begin{equation}
  \label{eq:fpiWS}
  f_{\pi,\text{WS}}^2 \equiv f_\pi^2 \frac{K_0(r)}{K_0(0)}
\end{equation}
as an in-medium, $r$-dependent pion decay ``constant''.

On the other hand, the non-local contribution 
$\widetilde{E}^{(2)}_{\text{WS}}$, --- the in-medium extension of 
(\ref{eq:enl}), ---  can be cast into the following form:
\begin{eqnarray}
  \widetilde{E}^{(2)}_{\text{WS}} &=& \frac{8\pi N_c M}{f_\pi^2}
    \sum_{\alpha\alpha'}\big\{ \frac{1}{3}
    f_0(\alpha_{0}) \Delta K_0(\alpha_{0},\alpha'_{0}) f_0(\alpha'_{0}) 
    + f_1(\alpha_{1}) \Delta K_1(\alpha_{1},\alpha'_{1}) 
    f_1(\alpha'_{1}) \nonumber \\
  && \qquad\qquad\qquad + \frac{2}{3}
    f_2(\alpha_{2}) \Delta K_2(\alpha_{2},\alpha'_{2}) f_2(\alpha'_{2}) 
    \big\} ,
\end{eqnarray}
where
\begin{mathletters}
\begin{eqnarray}
  f_0(\alpha_{0}) &=& M^{1/2} \int_0^R dr\,r^2 \rho_{\alpha_{0}}(r)
    \left[ \cos\theta(r)\theta'(r)+2\frac{\sin\theta(r)}{r} \right] \\
  \label{eq:f1}
  f_1(\alpha_{1}) &=& M^{1/2} \int_0^R dr\,r^2 \rho_{\alpha_{1}}(r) 
    \sin\theta(r)\theta'(r) \\
  f_2(\alpha_{2}) &=& M^{1/2} \int_0^R dr\,r^2 \rho_{\alpha_{2}}(r)
    \left[ \cos\theta(r)\theta'(r)-\frac{\sin\theta(r)}{r} \right]
\end{eqnarray}
\end{mathletters}
and
\begin{mathletters}
\begin{eqnarray}
  \Delta K_{l}(\alpha_{l},\alpha'_{l}) &=& \frac{1}{8\pi^2}
    \sum_{LL'}(2L+1)(2L'+1)\left( \begin{array}{ccc}
                                   L&L'&l\\ 
                                   0&0&0
                                   \end{array} \right)^2
    \kappa_{lLL'}(\alpha_{l},\alpha'_{l}) \\
  \kappa_{lLL'}(\alpha_{l},\alpha'_{l}) &=& \sum_{\alpha_{L}\alpha_{L'}}
    \frac{\pi}{E_{\alpha_{L}}} \left[ 
    \frac{C\left[\left(\alpha_{L}+\alpha_{L'}\right)/2\right]}{E_{\alpha_{L'}}
    (E_{\alpha_{L}}+E_{\alpha_{L'}})} - \frac{C(\alpha_{L})}{2E^2_{\alpha_{L}}}
    \right] \xi(\alpha_{l},\alpha_{L},\alpha_{L'}) 
    \xi(\alpha'_{l},\alpha_{L},\alpha_{L'}) \\
  \xi(\alpha_{l},\alpha_{L},\alpha_{L'}) &=& \int_0^R dr\,r^2
    \rho_{\alpha_{l}}(r) \rho_{\alpha_{L}}(r) \rho_{\alpha_{L'}}(r).
\end{eqnarray}
\end{mathletters}

One should now notice that the straightforward application of (\ref{eq:Seff2})
to the Wigner-Seitz cell is incomplete, because it does not account for the
Casimir energy intrinsically connected with the change of topology, which is 
present even in the absence of background fields. As a matter of fact one 
should write:
\begin{eqnarray}
\label{eq:Seff3}
    S_{\text{eff}}[U] &=& -\frac{i}{2} N_c\, \text{tr}\, 
    \left\{ \log\left[\Box + M^2+i M \rlap/\partial U_5(x)\right]_{\text{WS}} -
            \log\left[\Box + M^2 \right]_{\text{free}} \right\} \nonumber \\
                    &\equiv& -\frac{i}{2} N_c\, \text{tr}\, 
            \log\left[\frac{\Box + M^2+i M \rlap/\partial 
            U_5(x)}{\Box + M^2}\right]_{\text{WS}} +  
            \Delta S_{\text{eff}}^{\text{Casimir}} ,
\end{eqnarray}
where
\begin{eqnarray}
    \Delta S_{\text{eff}}^{\text{Casimir}} &\equiv& -\frac{i}{2} N_c\, 
      \text{tr}\, 
            \left\{ \log\left[\Box + M^2 \right]_{\text{WS}}
            -  \log\left[\Box + M^2 \right]_{\text{free}}  \right\} .
\label{eq:Scasimir}
\end{eqnarray}
Note that the first term in (\ref{eq:Seff3}) is just the application of
Eq. (\ref{eq:Seff2}) to the Wigner-Seitz case; on the other hand, the second 
term is the genuine Casimir energy due to the change of configuration space. 
By performing the intermediate algebra one obtains
\begin{eqnarray}
   \Delta S_{\text{eff}}^{\text{Casimir}} &\equiv& 4 N_c T 
                                \left\{ \sum_l \sum_{\alpha_l}
                                  (2 l+1) \sqrt{M^2+ \frac{\alpha_l^2}{R^2}} -
                        \lim_{R\rightarrow \infty} \sum_l \sum_{\alpha_l}
                        (2 l+1) \sqrt{M^2+ \frac{\alpha_l^2}{R^2}} \right\} .
\label{eq:Scasimir1}
\end{eqnarray}
The Casimir energy, which is obtained by dividing this expression by the time 
$-T$, has now the expected form: It is the difference between the zero-point 
energy in the finite volume, obtained by filling all the negative energy 
orbitals in the Dirac sea, and the same expression in free space. 
As it stands, however, Eq. (\ref{eq:Scasimir1}) is badly divergent and needs to
be regularized. Such a task has been recently carried out for massive fermions,
in the context of the MIT bag model, in Ref.~\cite{Eli98} by means of the zeta
function regularization technique. In the present work we will neglect this 
contribution to the total energy.

\subsection{ Regularization of integrals and sums }
\label{subsec:regular}

In calculating the vacuum contributions to the physical observables one has
to deal with the appearance of divergent expressions. In this paper we consider
two different regularization schemes, applying a momentum cutoff and using the
Pauli-Villars regularization. 

In the first case we introduce a regulating function, which suppresses the 
contribution to integrals and sums at momenta $k \gg \Lambda$, where 
the scale $\Lambda$ is determined by fitting the pion decay constant in free
space (see Eq.~(\ref{eq:K(0)}). The regulating function we have chosen has the 
form
\begin{equation}
  C(k) = \frac{1+ 1/e}{\exp\left[\left(k^2-\Lambda^2\right)/\Lambda^2\right]+1}
  .
\end{equation}

In the Pauli-Villars regulating scheme \cite{Kub99}, on the other hand, the 
divergent contributions are eliminated through the subtraction
\begin{equation}
  K(x,x') \rightarrow K(x,x') - K^{\text{PV}}(x,x') ,
\end{equation}
where $K(x,x')$ is the propagator previously defined, while in 
$K^{\text{PV}}(x,x')$ the quark mass $M$ has been substituted by the mass scale
$M_{\text{PV}}$, obtained again by fitting the free space pion decay constant:
\begin{equation}
  \label{eq:MPV}
  M_{\text{PV}} = M \exp\left(\frac{2 \pi^2 f_\pi^2}{N_c M^2}\right) .
\end{equation}

\subsection{ Pion decay constant }
\label{subsec:fpi}

Let's now consider the axial current. Its valence part is 
\begin{equation}
  A_a^\mu(x) = \overline{\psi}(x) \frac{\tau^a}{2} \gamma^\mu\gamma_5 \psi(x) .
\end{equation}
We wish to calculate also the vacuum (i.e. from the Dirac sea at the 
one-quark-loop level) contribution to the axial current: This can be done by 
defining the generating functional
\begin{equation}
  W[a] = \int \left[ d\overline{\psi} d\psi \right] 
    \exp\left\{ i \int d^4x \left[ \overline{\psi} \
    \left( i \rlap/\partial-M U_5 \right) 
    \psi - a_\mu^a A_a^\mu \right] \right\} ,
\end{equation}
where $a_a^\mu$ are classical axial sources coupled to the quantum axial
current; then, the vacuum axial current can be obtained by means of a 
functional derivative with respect to the source as
\begin{equation}
  A_{a,\text{vac}}^\mu (x) = i \frac{\delta}{\delta a_\mu^a(x)} \ln W[a]
    \Bigg|_{a^a_\mu=0} .
\end{equation}
Calculation provides
\begin{equation}
  A_{a,\text{vac}}^\mu (x) = \frac{4 N_c M^2}{f_\pi^2} \int d^4x' K(x,x') 
    \left[ \partial^\mu \phi_0(x') \phi^a(x) - 
    \partial^\mu \phi^a(x') \phi_0(x) \right] .
\end{equation}
By setting
\begin{equation}
  K(x,x') = K_0(r) \delta^4(x-x') + \Delta K(x,x') ,
\end{equation}
where $K_0(r)$ is given by Eq.~(\ref{eq:Kr}), one is able to  write
\begin{eqnarray}
  \label{eq:AmuWS}
  A_a^\mu (x) &=& \overline{\psi}(x) \frac{\tau^a}{2} \gamma^\mu\gamma_5 
    \psi(x) + \frac{4 N_c M^2}{f_\pi^2} K_0(r) 
    \left[ \partial^\mu \phi_0(x) \phi^a(x) - 
    \partial^\mu \phi^a(x) \phi_0(x) \right] \nonumber \\
  && \quad + \frac{4 N_c M^2}{f_\pi^2} \int d^4x' \Delta K(x,x') 
    \left[ \partial^\mu \phi_0(x') \phi^a(x) - 
    \partial^\mu \phi^a(x') \phi_0(x) \right] ,
\end{eqnarray}
in which local and non-local contributions have been separated.

Remembering now that the pion decay constant is defined as
\begin{equation}
  \langle 0 | A_a^\mu(x) | \pi^b(p) \rangle = - i p^\mu f_\pi \delta_{a,b} 
  e^{-i p\cdot x} ,
\end{equation}
one is able to obtain the in-medium pion decay constant as
\begin{equation}
  f_{\pi,\text{WS}}^2(r) = 4 N_c M^2 K_0(r) ,
\end{equation}
that is expression (\ref{eq:fpiWS}), which depends, in a medium, on the radial
coordinate.

An estimate of the average value of $f_\pi$ at fixed density can be obtained
by calculating the constant value of $f_{\pi,\text{WS}}$ that would yield the
same pion kinetic energy as the $r$-dependent one (compare 
Eqs. ~(\ref{eq:Ekinfree}) and ~(\ref{eq:EkinWS})):
\begin{equation}
  \langle f_\pi \rangle^2 = \frac{\int_0^R dr\, f_{\pi,\text{WS}}^2(r) 
    \left[ r^2{\theta'}^2(r) + \sin^2\theta(r) \right]}{\int_0^R dr
    \left[ r^2{\theta'}^2(r) + \sin^2\theta(r) \right]} .
\label{eq:fpiave}
\end{equation}

\subsection{ Axial coupling constant }
\label{subsec:gA}

The axial coupling constant, for a system with a finite pion mass is given by
\begin{equation}
  \label{eq:gA}
  \frac{1}{2} g_A = \langle p \uparrow | \int d\bbox{r} A_3^z | p \uparrow 
    \rangle .
\end{equation}
The matrix element between proton states requires one to adopt a collective
quantization procedure in order to project out the correct quantum numbers.
In the case of $g_A$, --- as discussed, e.~g., in Ref.~\cite{Bro86}, --- it
amounts simply to multiply the expression for the hedgehog axial current by the
matrix element of the cranking operator, $-1/3$. Furthermore, as discussed in
Ref.~\cite{Adk83}, the $m_\pi=0$ limit requires one to perform first the 
angular integral and then the radial integral in (\ref{eq:gA}), multiplying the
result by a factor $3/2$.

Use of Eq.~(\ref{eq:AmuWS}) yields
\begin{equation}
  g_A = \int_0^R dr\,r^2 \left[\frac{N_c}{2}(u^2-\frac{1}{3}v^2) 
    - \frac{4\pi}{3} f_{\pi,\text{WS}}^2 \left(\theta'
    + \frac{\sin(2\theta)}{r} \right) \right]
    - \int d\bbox{r} \Delta A_3^z  ,
\end{equation}
with
\begin{eqnarray}
 \int d\bbox{r} \Delta A_3^z  &=& -\frac{16 \pi N_c M^2}{3} 
    \sum_{\alpha,\alpha'}\left\{ \overline{f}_0(\alpha_0) 
    \Delta K_0(\alpha_0,\alpha'_0) f_0(\alpha'_0) + \overline{f}_1(\alpha_1) 
    \Delta K_1(\alpha_1,\alpha'_1) f_1(\alpha'_1) \right\} 
\nonumber 
\end{eqnarray}
and
\begin{eqnarray}
  \overline{f}_0(\alpha_0) &\equiv& M^{3/2} \int_0^R dr \ r^2 \ 
    \rho_{\alpha_0}(r) \left( \cos\theta(r)-1\right) \nonumber \\
  \overline{f}_1(\alpha_1) &\equiv& M^{3/2} \int_0^R dr \ r^2 \ 
    \rho_{\alpha_1}(r) \sin\theta(r) . \nonumber 
\end{eqnarray}
These are the expressions we shall employ in the next Section to evaluate the 
in-medium modification of the axial coupling constant.

\subsection{ Equations of motion }
\label{subsec:EoM}

The equations of motion are found by minimizing the total energy
(\ref{eq:Etot}) with respect to the Dirac fields, $u$ and $v$, and to the
chiral angle, $\theta$. It is convenient to write them in a dimensionless form
by letting $q\to Mq$ and introducing 
\begin{equation}
  x=Mr, \quad \tilde{u}(x)=M^{-3/2}u(r), \quad \tilde{v}(x)=M^{-3/2}v(r) .
\end{equation}
Minimization of $\epsilon_{\text{tot}}\equiv E_{\text{tot}}/M=
\epsilon_{\text{val}}+\epsilon^{(2)}_{\text{vac}}$ then yields
\begin{mathletters}
\begin{eqnarray}
  \frac{d\tilde{u}}{dx} &=& -\sin\theta\,\tilde{u} - (\epsilon_{\text{val}} 
    + \cos\theta)\tilde{v} \\
  \frac{d\tilde{v}}{dx} &=& -\left(\frac{2}{x}-\sin\theta\right)\tilde{v} + 
    (\epsilon_{\text{val}} - \cos\theta)\tilde{u} \\
  \frac{d^2\theta}{dx^2} &+& \frac{2}{x}\frac{d\theta}{dx} -
    \frac{\sin 2\theta}{x^2} + \frac{N_c M^2}{4\pi f_\pi^2} 
    [(\tilde{u}^2-\tilde{v}^2)\sin\theta + 2\tilde{u}\tilde{v}\cos\theta] 
    \nonumber \\
  &+& \frac{2}{\pi}\sin\theta\int_0^\infty dq\,q^2
    \left[\frac{K(q)}{K(0)}-1\right] j_0(qx) {\cal C}(q) 
    \nonumber \\
  &-& \frac{2}{\pi}\cos\theta\int_0^\infty dq\,q^2
    \left[\frac{K(q)}{K(0)}-1\right] j_1(qx) {\cal S}(q) = 0 ,
\end{eqnarray}
\end{mathletters}
where
\begin{mathletters}
\begin{eqnarray}
  {\cal C}(q) &=& \int_0^\infty dx(qx)^2 j_0(qx)[\cos\theta(x)-1] \\
  {\cal S}(q) &=& \int_0^\infty dx(qx)^2 j_1(qx)\sin\theta(x) ,
\end{eqnarray}
\end{mathletters}
and $K(q)$ is still given by (\ref{eq:Kq}), but now $k$, $k'$ are dimensionless
and $E_k\to\tilde{E}_k=\sqrt{k^2+1}$.
This is a set of integro-differential equations that has to be solved
iteratively as shown, for instance, in Ref.~\cite{Adj92}.

The same procedure, applied to the system enclosed into the WS cell, yields
\begin{mathletters}
\label{eq:equations}
\begin{eqnarray}
  \label{eq:eqdirac1}
  \frac{d\tilde{u}}{dx} &=& -\sin\theta\,\tilde{u} - (\epsilon_{\text{val}} 
    + \cos\theta)\tilde{v} \\
  \label{eq:eqdirac2}
  \frac{d\tilde{v}}{dx} &=& -\left(\frac{2}{x}-\sin\theta\right)\tilde{v} + 
    (\epsilon_{\text{val}} - \cos\theta)\tilde{u} \\
  \label{eq:eqthetaWS}
  \frac{d^2\theta}{dx^2} &+& 2\left[\frac{1}{x}+\frac{d}{dx}\ln K_0(x)\right]
    \frac{d\theta}{dx} -
    \frac{\sin 2\theta}{x^2} + \frac{1}{16 \pi K_0(x)} 
    [(\tilde{u}^2-\tilde{v}^2)\sin\theta + 2\tilde{u}\tilde{v}\cos\theta] 
    \nonumber \\
  && \qquad + \sin\theta\frac{\widetilde{W}_a(x)}{K_0(x)} 
            + \cos\theta\frac{\widetilde{W}_b(x)}{K_0(x)}=0 ,
\end{eqnarray}
\end{mathletters}
where $K_0(x)$ is still given by (\ref{eq:Kr}), but now
$E_{\alpha_{l}/R}\to\tilde{E}_{\alpha_{l}/R}=\sqrt{(\alpha_{l}/X)^2+1}$ and
$\rho_{\alpha_{l}}(r)\to\tilde{\rho}_{\alpha_{l}}(x)=
M^{-3/2}\kappa_{\alpha_{l}}j_{l}(\alpha_{l}x/X)$, having set $X=MR$.
In (\ref{eq:eqthetaWS}) we have also set
\begin{mathletters}
\begin{eqnarray}
  \widetilde{W}_a(x) &=& \sum_{\alpha_{1}\alpha'_{1}}
    \left[\frac{2}{x}\tilde{\rho}_{\alpha_{1}}(x)
    +\frac{d\tilde{\rho}_{\alpha_{1}}(x)}{dx}\right]
    \Delta K_1(\alpha_{1},\alpha'_{1}) f_{1}(\alpha'_{1}) \\
  \widetilde{W}_b(x) &=& \frac{1}{3} \sum_{\alpha\alpha'} \left\{
    \frac{d\tilde{\rho}_{\alpha_{0}}(x)}{dx}
    \Delta K_0(\alpha_{0},\alpha'_{0}) f_{0}(\alpha'_{0})
    +2\left[\frac{3}{x}\tilde{\rho}_{\alpha_{2}}(x)
    +\frac{d\tilde{\rho}_{\alpha_{2}}(x)}{dx}\right]
    \Delta K_2(\alpha_{2},\alpha'_{2}) f_{2}(\alpha'_{2}) \right\}. 
    \nonumber \\
\end{eqnarray}
\end{mathletters}

\subsection{ Boundary conditions }
\label{subsec:boundary}

In the free space problem the boundary conditions on the fields are determined
straightforwardly. From inspection of the Dirac equations one has 
$\tilde{v}(0)=0$ and $\{\tilde{u}(x),\tilde{v}(x)\}\to0$, $x\to\infty$;
finiteness of the energy requires $\theta(x)\to0$ when $x\to\infty$, whereas by
choosing $\theta(0)=\pi$ one fixes to unity the topological charge associated
to the pion field. This is sometimes interpreted as the baryon number, but not
in the present model \cite{Dia88}, where it is connected to the number of
valence levels that are pushed out of the Dirac sea. Baryon number is fixed by
the normalization condition $\int_0^\infty dx x^2\tilde{\rho}(x)=1$, where 
$\tilde{\rho}(x)=\tilde{u}^2(x)+\tilde{v}^2(x)$ represents the (dimensionless)
baryon density.

If we consider now nuclear matter as a collection of hedgehog field
configurations centered at lattice points, we have to impose periodic boundary
conditions on the fields (Bloch's theorem). One then obtains a band structure,
that is a continuous set of states and an energy gap above the highest energy
state. The physical meaning of a band structure of the quark levels is not
clear in the present context, since it implies long-range correlations among
quarks and one should be aware that this (and similar) model does not account
for confinement. We shall come back to this point later.

In order to simplify the approach, the Wigner-Seitz approximation assumes that
the cell be symmetric: Then, the state at the bottom of the band is also
spherically symmetric and the wave functions are flat. To describe the other
states, --- and, in particular, the top of the band, --- many different
assumptions have been made in the literature (see, e.~g., Ref.~\cite{Web98} for
a brief summary).
In the calculations presented in Sect.~\ref{sec:res}, --- besides the
normalization condition $\int_0^X dx x^2\tilde{\rho}(x)=1$, which fixes the
baryon number, --- we shall use three distinct sets of boundary conditions 
(sets I, II and III). 

In order to make contact with previous calculations, we follow for set I the
choice of Ref.~\cite{Gle86}, where the authors insist in maintaining unit
topological charge inside the cell:
\begin{eqnarray}
  \label{eq:bcI}
  \theta(0) = \pi, \quad     \theta(X) &=& 0 \nonumber \\
  \tilde{v}(0) = 0, \quad \tilde{v}(X) &=& 0 \quad 
    \text{(``bottom'' of the band)} \\
    \tilde{u}(X) &=& 0 \quad \text{(``top'' of the band)}, \nonumber
\end{eqnarray}
where $\tilde{v}(X)=0$ which implies $\tilde{u}'(X)=0$.

For set II we choose to impose ``flatness'' also on the chiral angle, as in
Refs.~\cite{Nym70,Amo98}. Since in general one has $\theta(X)\ne0$ at the
boundary, from inspection of the Dirac equations one sees that 
$\tilde{v}(X)=0$ no longer implies $\tilde{u}'(X)=0$. We have chosen to 
impose the physically motivated constraint
\begin{equation}
  \label{eq:bcII}
  \label{eq:rho'}
  \tilde{\rho}'(X)=0 ,
\end{equation}
that is we require the {\em flatness of the baryon density} at the cell 
boundary.
Set II then turns out to be
\begin{eqnarray}
  \theta(0) = \pi, \quad     \theta'(X) &=& 0 \nonumber \\
  \tilde{v}(0) = 0, \quad  \tilde{v}(X) &=& 
    \frac{X\cos\theta(X)\mp\sqrt{X^2-2X\sin\theta(X)}}{X\sin\theta(X)-2}
    \tilde{u}(X) \quad  \text{(``bottom''/``top'')} .
\end{eqnarray}
For set III we require the flatness of the baryon density together with the
requirement of unit topological charge:
\begin{eqnarray}
  \label{eq:bcIII}
  \theta(0) &=& \pi, \quad     \theta(X) = 0 \nonumber \\
  \tilde{v}(0) &=& 0, \quad  
    \left\{ \begin{array}{l}
              \tilde{v}(X)=0 \\
              \tilde{v}(X)=-X\tilde{u}(X)
            \end{array} \right. \quad
            \begin{array}{l}
              \text{(``bottom'')} \\
              \text{(``top'')}
            \end{array} .
\end{eqnarray}
One might be tempted to interpret the two solutions of the equation 
(\ref{eq:rho'}) as corresponding to the bottom and top of the energy band. 
Indeed, from set III one sees that $\tilde{v}(X) = 0$ corresponds to the bottom
of the band in set I, whereas in the non-relativistic limit ($\tilde{v}\to0$)
the second condition in (\ref{eq:bcIII}), that is 
$-X\tilde{u}(X)=\tilde{v}(X)$, reduces to $\tilde{u}(X) = 0$ as in set I.
 
A word of caution is necessary in analyzing the boundary 
conditions. In our view one cannot accept without question the presence of
a band of quark states. The presence of a band, in fact, would be affected by
confinement, which is absent in the present model.
In free space, where the quarks are deeply bound in the ground state of 
the chiral fields, this shortcoming is not crucial (of course, the study of 
the highly excited states of the nucleon would then be problematic).  
In the medium, however, because of the lack of confinement, quarks exhibit 
unrealistic long--range correlations.  
As a result, one observes a relatively large  probability of having a quark 
sitting at the surface of the WS cell. 

In the next Section we shall see that when the density of the medium increases 
the quark density tends to be more concentrated in the interior of the bag for
the {\em lowest} end of the energy band (lowest when all effects have been 
included), whereas the {\em opposite} happens on the {\em upper} end.
As a consequence, quarks sitting in the upper part of the energy band would
be more affected by the confining forces than the ones in the lower part.
Since the confining forces would tend to reduce the quark density at the 
boundary, the net result would then be mainly a lowering, --- and hence a 
narrowing, ---of the highest end of the band.

Another problem one should cope with in identifying the top and bottom energy 
levels is posed by the presence of the spurious center of mass energy. Although
this contribution has been neglected in most existing calculations of in-medium
properties in the Wigner-Seitz approximation, it turns out not only to
be sizeable but also to affect the relative position of the ``top'' and 
``bottom'' levels, as it will be discussed in the next section.

In the following, for definiteness, we shall follow the nomenclature adopted 
in the literature and we shall label ``bottom'' and ``top'' the two solutions 
of each set of boundary conditions, as indicated in (\ref{eq:bcI}), 
(\ref{eq:bcII}) and (\ref{eq:bcIII}).

\section{ Results }
\label{sec:res}

The numerical results have been obtained by integrating the equations of 
motion (\ref{eq:equations}) for the quarks and the meson fields, using an 
iterative procedure as, e.g., in Ref.~\cite{Adj92} and moving from larger 
to smaller values of $R$. 
For a given value of $R$, i.e. for a given density, a self-consistent 
solution has been found by using as initial ansatz the self-consistent 
chiral profile obtained at the previous value of $R$. 
The non-local term in Eq.(\ref{eq:eqthetaWS}) has been switched on 
adiabatically in order to allow a better convergence\footnote{The sums over the
modes in the orthonormal and complete basis inside the spherical cell have 
been restricted to $l\le 15$ and to the first 30 roots $\alpha_l$, which
provides a good degree of accuracy.}.

At the smallest density, corresponding to $R = 5$ fm, an exponential 
profile $\theta(r) = \pi \ exp(-r/r_0)$ has been used as initial ansatz.
A dynamical (constituent) quark mass $M = 350$ MeV has been 
assumed in the calculations. This is a value suggested by the phenomenology
of the single nucleon \cite{Dia88,Dia89,Wak91}, which also turns out to be in 
the range ($300-400$ MeV) where the second order expansion for the effective 
action works well \cite{Adj92}. 

The divergences that appear both in free space (in the momentum integrals) and 
in the medium (in the sums) have been regulated as explained in  Sect. 
\ref{subsec:regular}, using both a regulating function and the Pauli-Villars 
regularizations, whose parameters have been fixed by fitting the free space 
value of the pion decay constant.
For $M=350$ MeV, the cutoff in the regulating function turns out to be
$\Lambda\cong500$ MeV, whereas the mass scale $M_{\text{PV}}$ of the
Pauli-Villars regularization is given by Eq.~(\ref{eq:MPV}).
The two regularization schemes yield qualitatively similar results and in the
following we shall display only the outcome from the regulating function
approach. 

Before discussing the results a comment on the flat basis we have adopted is 
in order. In fact, this basis contains a zero momentum state, that is, a term 
constant in space and proportional to $R^{-3/2}$. When $R \rightarrow 0$,
it would give rise to divergences. As explained in Appendix \ref{app:basis},
the appearance of such a mode is peculiar of the flat basis: One might 
introduce bases infinitesimally close to the flat one, in which the zero mode 
is absent and its strength is distributed among all the other modes. 
In these alternative bases the effect of the zero mode would be given by an 
infinite sum of infinitesimal contributions; since we are regularizing the 
sums, only a finite number of modes enters into the calculation of physical 
quantities and the contribution to them of the redistributed strength of the 
zero mode is infinitesimal. Hence, we can cure the divergence simply by 
dropping the zero mode.

\begin{figure}[p]
\begin{center}
\mbox{\epsfig{file=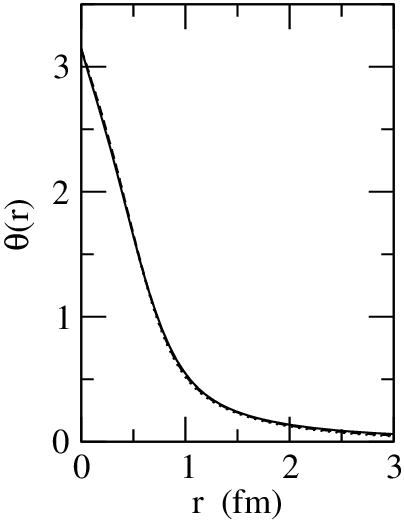,height=8cm}}
\caption{The chiral angle in free space (solid line) and in the WS 
approximation at $R=5$ fm, using the boundary conditions of set II (dashed) and
of sets I and III (dotted). }
\label{fig:thetafree}
\vskip 1cm
\mbox{\epsfig{file=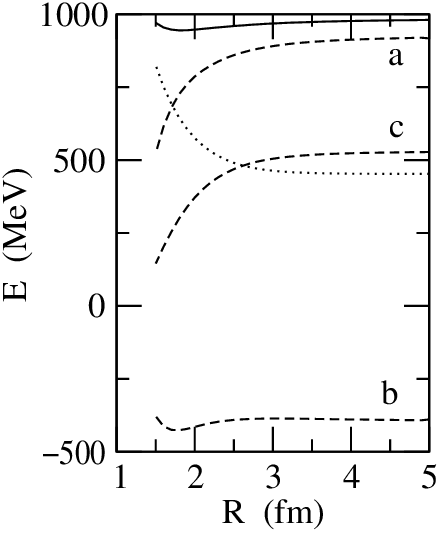,height=8cm}}
\caption{ Total energy of the WS bag (solid) for the ``bottom'' solution of set
I. Also shown are the valence (dotted) and vacuum (dashed c) contributions; of
the latter one, we display also the local (dashed a) and non-local (dashed b)
components. }
\label{fig:glenn}
\end{center}
\end{figure}
In order to exhibit the convergence of the WS calculations to the free space 
results, let us start by comparing, in Fig.~\ref{fig:thetafree}, the chiral 
angle obtained by solving the equations of motion in free space (solid lines) 
to the WS 
solutions corresponding to $R=5$ fm and obeying the boundary conditions of 
set II (dashed line) and of sets I and III (dotted line).
Note that at such a low density, the top and bottom energy levels practically
coincide. 
In both cases a regulating function has been used to regulate sums and 
integrals. 
As expected, the difference between the solutions inside the WS cell and 
the one in free space is barely  noticeable at this density. 
By looking at Table \ref{tab:table1} we also notice that the free space 
energies are recovered with good precision, using any of the boundary 
conditions.
\begin{table}
\caption{Comparison between the energies (in MeV) obtained in free space
and in the WS cell using the boundary conditions of sets I-III for $R=5$ fm.}
\begin{tabular}{c|c|ccc|c}
  & {$E_{\text{val}}$} & \multicolumn{3}{c|}{$E^{(2)}_{\text{vac}}$} & 
  {$E_{\text{tot}}$} \\
  &             & {$E_{\text{kin}}$} & {$\widetilde{E}^{(2)}$} 
  & {$E_{\text{kin}} + \widetilde{E}^{(2)}$} & \\
\tableline
free    & 452.4 & 922.4 & -387.6  & 534.8 & 987.2  \\
II   & 443.6 & 921.8 & -386.2  & 535.6 & 979.2 \\
I and III & 452.9 & 917.5 & -389.6  & 527.9 & 980.8  \\
\end{tabular}
\label{tab:table1}
\end{table}

\begin{figure}[p]
\begin{center}
\mbox{\epsfig{file=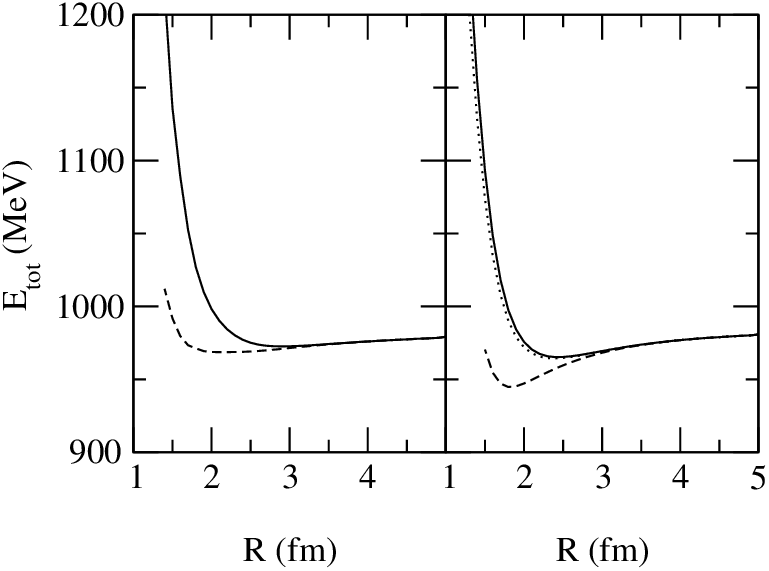,height=8cm}}
\caption{ Total energy of the WS bag. The left panel corresponds to the
boundary conditions of set II (solid and dashed lines for the ``top'' and 
``bottom'' solutions, respectively), whereas the right panel corresponds to the
boundary conditions for the ``bottom'' solutions of sets I and III (dashed) and
for the ``top'' solutions of sets I (dotted) and III (solid). }
\label{fig:etotreg}
\vskip 1cm
\mbox{\epsfig{file=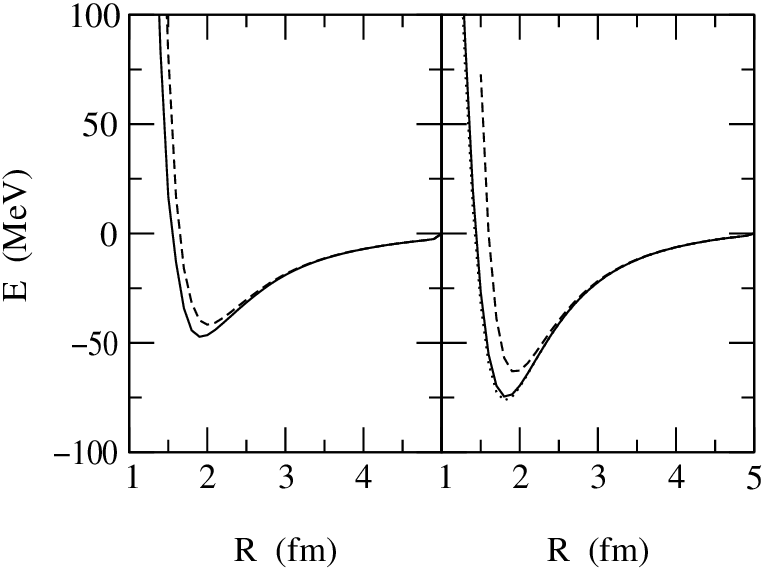,height=8cm}}
\caption{ Binding energy obtained by taking out the spurious center-of-mass 
energy contribution and subtracting the energy at the lowest density (here 
$R=5$ fm). The left panel corresponds to the boundary conditions of set II 
(solid and dashed lines for the ``top'' and ``bottom'' solutions, 
respectively), whereas the right panel corresponds to the boundary conditions 
for the ``bottom'' solutions of sets I and III (dashed) and for the ``top'' 
solutions of sets I (dotted) and III (solid). }
\label{fig:bereg}
\end{center}
\end{figure}

An important difference between our work and previous calculations employing 
chiral quark models \cite{Web98,Ban85,Gle86,Hah87} is due to the inclusion in 
our calculations of non-local effects stemming from the vacuum contribution (at
the one-quark-loop level). In Fig.~\ref{fig:glenn} we display, as a function of
$R$, the WS cell total energy (solid line), separated in the valence (dotted
line) and vacuum (dashed c line) terms, for the ``bottom'' solution of set I.
Also shown are the separated local, --- that is, kinetic, --- (dashed a line)
and non-local (dashed b line) contributions.
The local term displays a behavior similar to the results of, e.~g.,
Ref.~\cite{Hah87}: 
Actually, in the chiral quark model employed in that paper, the kinetic meson
contribution is present at the classical level, whereas in our case it is
dynamically generated from the vacuum. This implies, as we saw in
Sect.~\ref{sec:model}, a $r$-dependent pion decay constant (see 
Eqs.~(\ref{eq:EkinWS}) and (\ref{eq:fpiWS})); We have however checked that 
setting $f_{\pi,\text{WS}}$ constant in our calculation, the results of
Ref.~\cite{Hah87} are recovered.
On the other hand, the non-local vacuum contribution provides substantial 
attraction and displays a moderate dependence on $R$; however, it turns out 
that the valence and kinetic meson terms compensate each other to a large
extent yielding a total contribution without any minimum and the non-local
term is then instrumental in order to get the (rather shallow) minimum 
displayed by the solid line. 

By looking at Fig.~\ref{fig:etotreg} one notices that solutions to the 
equations of motion are no longer found below $R \approx 1.4$ fm. 
This depends also on the choice of the boundary conditions and in the other
cases discussed below we shall see that solutions are found till 
$R \approx 1.1$ fm. In Ref.~\cite{Hah87} solutions have been found 
till much higher densities ($R \approx 0.4$ fm). In the case of Fig.~2 in that 
paper, --- where a simple Lagrangian containing only terms up to second order 
in the pion field is used, --- this is due to the larger value for the 
constituent quark mass $M$ chosen in that work (in their notation 
$M = g f_{\pi} \approx 550$ MeV, where $g$ is the quark-meson coupling 
constant). We have checked that by dropping the non-local term and by
increasing $M$ their results can be recovered.
Indeed, from Eq.~(\ref{eq:equations}) one sees that the equations
of motion depend on the combination $X = M R$: By increasing $M$, one can 
lower the minimum value of $R$. 
However, the authors of Ref.~\cite{Hah87} are able to find solutions at higher
densities by using more complex Lagrangians containing terms of higher order in
the pion field. Following that path in our model would imply going beyond the
two-point approximation to the effective action of Sect.~\ref{subsec:effact2}.

In Fig.~\ref{fig:etotreg} we display the WS cell total energy as a function of 
the cell radius, i.e. of the density. The boundary conditions of set II and of 
sets I and III have been used in the left and right panels, respectively. 
In Fig.~\ref{fig:etotreg} we notice the presence of a very shallow minimum 
around a density corresponding to $R\approx2$ fm; this minimum is deeper for 
the ``bottom'' solutions. 

The occurrence of saturation in nuclear matter cannot however be stated by 
simply looking at these figures, because here the spurious 
energy contribution due to the center-of-mass motion has been neglected. 
This, as a matter  of fact, is a well known problem associated to the mean 
field approximation. 
An estimate of this effect in the chiral quark soliton model has been obtained 
in \cite{Pob92}, including also the vacuum (mesonic) contributions to the 
center of mass motion. The findings in that paper, --- that valence terms 
dominate as long as $E_{val} \geq 0$, --- make us feel confident in retaining 
only the latter in our estimate. 
It reads
\begin{eqnarray}
E_{\text{CM}} &=& \frac{\langle P^2 \rangle}{2 E_{\text{tot}}} \nonumber \\
  &=& - \frac{N_c}{2 E_{\text{tot}}} \left\{ R^2 \rho'(R) -
  \int_0^R dr \left[ r^2 ( {u'}^2(r)+{v'}^2(r)^2) + 2 v^2(r) \right] \right\} .
\end{eqnarray}
Of course, the validity of this assumption at finite density can only be 
checked through an explicit calculation of the  vacuum terms, which  
is however beyond the scope of the present analysis. 

\begin{figure}[p]
\begin{center}
\mbox{\epsfig{file=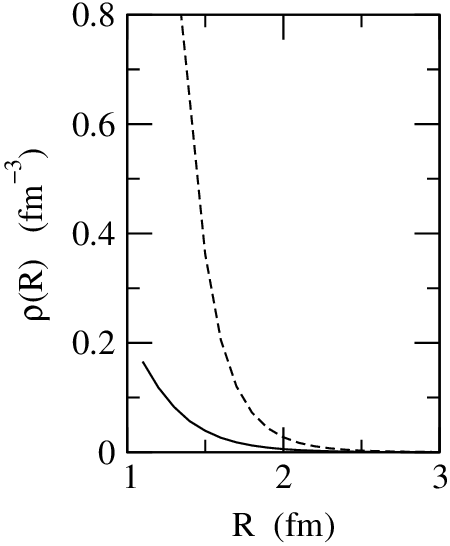,height=8cm}}
\caption{ Dependence of the baryon density upon the cell radius for the ``top''
(solid) and ``bottom'' (dashed) solutions of set II.}
\label{fig:rhotb}
\vskip 1cm
\mbox{\epsfig{file=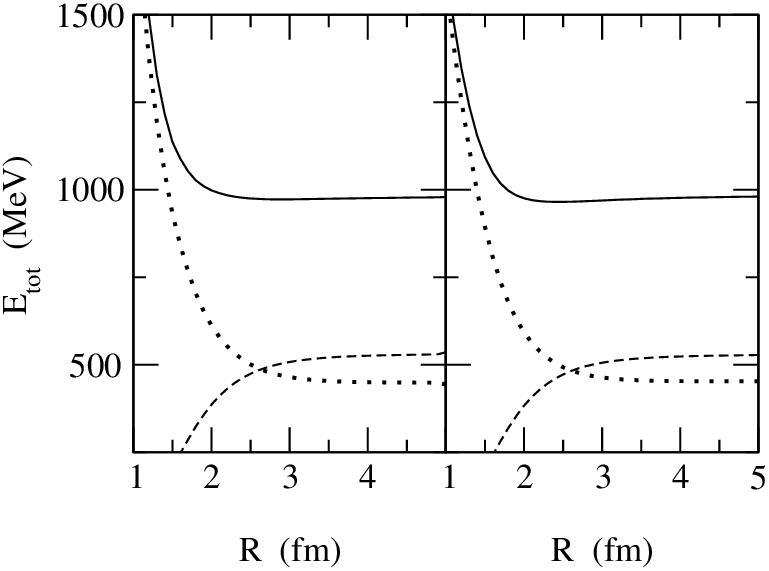,height=8cm}}
\caption{ Decomposition of the total energy (solid) into valence (dot) and 
vacuum (dashed) contributions, for the ``top'' solutions of set II (left panel)
and III (right panel). }
\label{fig:evalvac}
\end{center}
\end{figure}

In Fig.~\ref{fig:bereg} we plot the binding energy for the system taking out 
the spurious contributions stemming from the motion of the center of mass. 
In order to minimize the numerical uncertainty, the binding energy has been
obtained by subtracting the total energy at the lowest considered density 
($R = 5$ fm).
Interestingly, we find that, when the center-of-mass motion is taken out, 
a stronger minimum in the energy is found, roughly at the same density as 
in Fig.~\ref{fig:etotreg}, namely $R \approx 1.8$ fm. Moreover, the
boundary conditions of sets I and III provide more binding that those 
of set II.
Note that in nuclear matter one should have a binding energy of about 
$-16$ MeV at $R \cong 1.1$ fm.

We also notice  that the ``top'' solutions of the various sets of boundary
conditions provide more binding than the corresponding ``bottom'' solutions. 
The reason is easily understood by looking, for example, at 
Fig.~\ref{fig:rhotb}.

\begin{figure}[p]
\begin{center}
\mbox{\epsfig{file=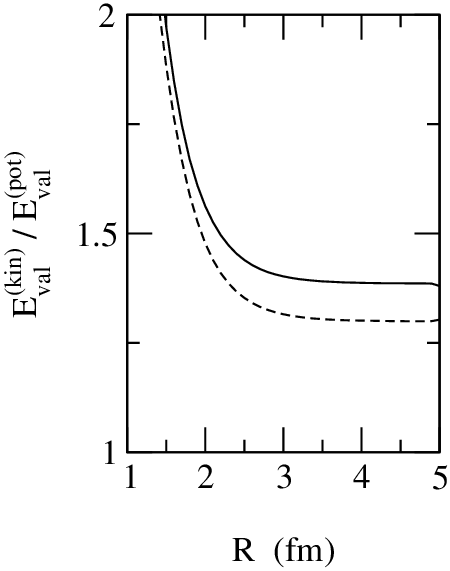,height=8cm}}
\caption{ The ratio between the kinetic and potential components of the valence
energy as a function of the cell radius, using the ``top'' boundary conditions
of set II (solid) and set III (dashed). }
\label{fig:ekinepot}
\vskip 1cm
\mbox{\epsfig{file=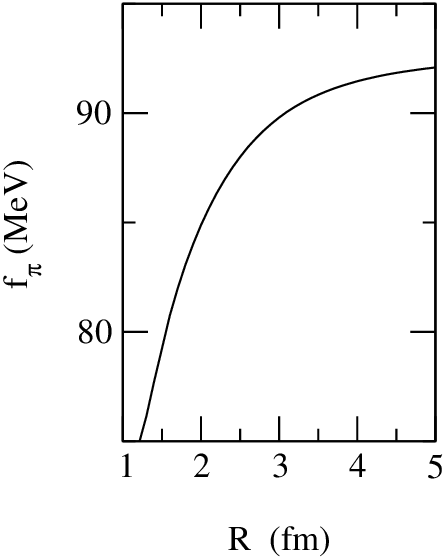,height=8cm}}
\caption{ Dependence of the pion decay constant on the the cell radius for the 
``top'' solution of set II. }
\label{fig:fpi}
\end{center}
\end{figure}

Here we plot the value of the baryon density at the surface of the cell,
i.e. $u^2(R)+v^2(R)$, as a function of the cell radius itself, for the boundary
conditions of set II (the solid and dashed lines corresponding to the top and 
bottom solutions respectively); sets I and III show a similar behavior. 
Since the top solution corresponds to a configuration in which the quarks 
are more ``compressed'' inside the cell, a larger kinetic energy, --- and 
therefore a larger center-of-mass motion, --- is associated with it.
This problem is of course absent in solid state physics, the original field of 
application of the Wigner-Seitz approximation, because the electron mass is 
indeed completely negligible with respect to the mass of the ions, which form 
the periodic structure. In the present case, even admitting the existence of a 
periodic structure in nuclear matter, the center-of-mass motion would be 
neglegible only in the large $N_c$ limit ($N_c \rightarrow \infty$), given the 
dependence of the total energy and of the center-of-mass energy on the number 
of colors, as $O(N_c)$ and $O(N_c^0)$ respectively. 
On the other hand, for $N_c=3$, the center-of-mass energy will vary, for each 
solution inside the band, by an amount comparable with the width of the band 
itself. Hence, the calculation of a reliable band structure  within this model 
is more delicate.

In Fig.~\ref{fig:evalvac} the total energy corresponding to the ``top'' 
solutions of set II (left) and III (right) is plotted, and decomposed into the 
valence (dotted) and vacuum (dashed) contributions. 
We observe that, at very low densities, the two components bear approximately 
the same strength, whereas at larger densities, below $R \cong 2$ fm, the 
valence contribution becomes dominant. 
It is also interesting to display, as a function of $R$, the ratio between 
the valence contributions that come from the quark kinetic term in the
Lagrangian and from the quark coupling to the mean field.
This is done in Fig.~\ref{fig:ekinepot}: As expected, this ratio increases 
with the density.

\begin{figure}[p]
\begin{center}
\mbox{\epsfig{file=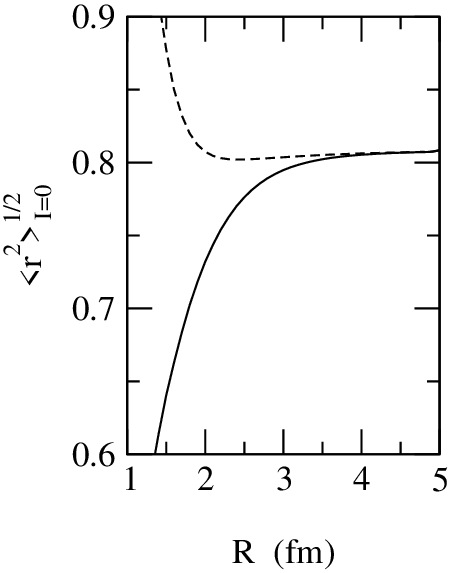,height=8cm}}
\caption{ Isoscalar mean square radius, as a function of the cell radius.
The solid line corresponds to the ``top'' solution of set II, whereas the 
dashed line corresponds to the ``bottom'' solution of Set III. }
\label{fig:sqrm}
\vskip 1cm
\mbox{\epsfig{file=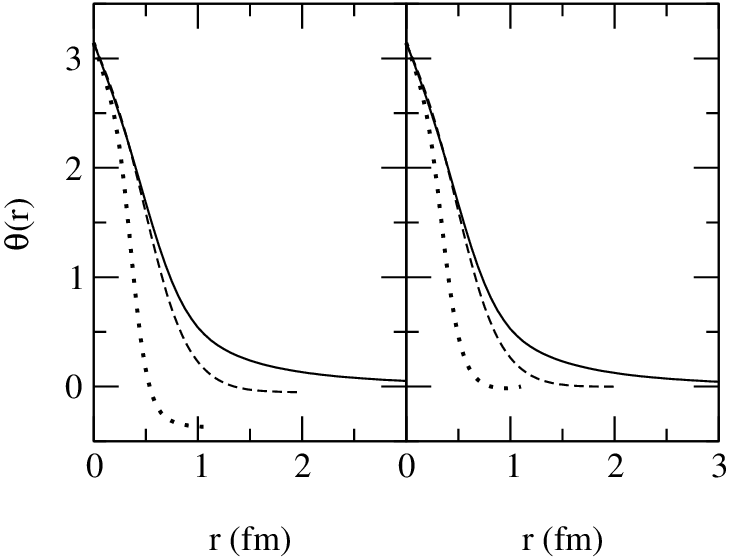,height=8cm}}
\caption{ Chiral angle as a function of the
distance from the center of the soliton for $R=5$ (solid), $2$ (dashed)
and $1.1$ fm (dotted), respectively. The left panel corresponds to the
``top'' solution of set II, whereas the right one corresponds to the ``top''
solution of set III. }
\label{fig:theta123}
\end{center}
\end{figure}

\begin{figure}[p]
\begin{center}
\mbox{\epsfig{file=fig_uv123.eps,height=8cm}}
\caption{ Large and small components of the Dirac spinor as a function of the
distance from the center of the soliton for $R=5$ (solid), $2$ (dashed)
and $1.1$ fm (dotted), respectively. The left panel corresponds to the
``top'' solution of set II, whereas the right one corresponds to the ``top''
solution of set III. }
\label{fig:uv123}
\vskip 1cm
\mbox{\epsfig{file=fig_rho123.eps,height=8cm}}
\caption{ Baryon density as a function of the distance from the center of the 
soliton for $R=5$ (solid), $2$ (dashed) and $1.1$ fm (dotted), respectively. 
The left panel corresponds to the ``top'' solution of set II, whereas the 
right one corresponds to the ``top'' solution of set III. }
\label{fig:rho123}
\end{center}
\end{figure}

\begin{figure}
\begin{center}
\mbox{\epsfig{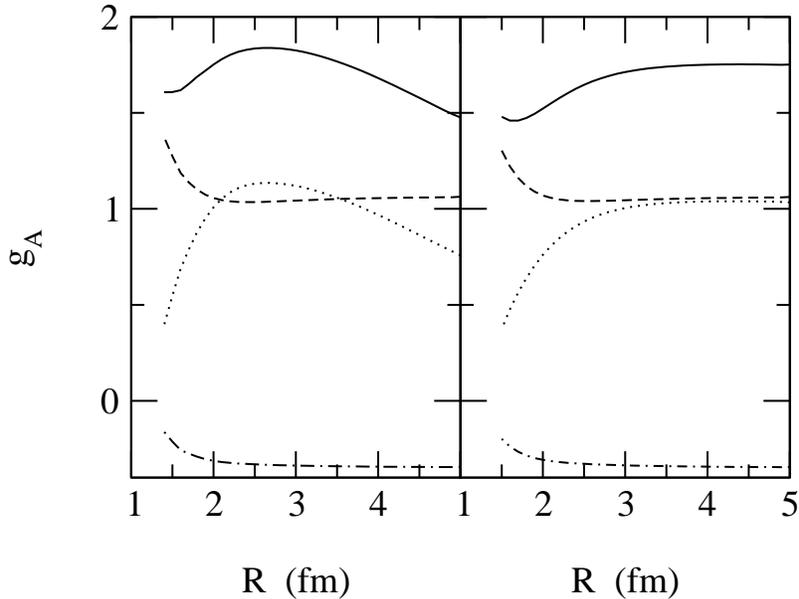}}
\caption{ The axial coupling constant as a function of $R$, for the ``top'' 
boundary conditions of set II (left panel) and set III (right panel).
The valence (dashed), local mesonic (dotted), non-local mesonic (dot-dashed) 
and total (solid) contributions are displayed. }
\label{fig:ga}
\end{center}
\end{figure}

In Fig.~\ref{fig:fpi} the average value of the pion decay constant in the cell,
defined in Eq. (\ref{eq:fpiave}), is plotted as a function of the cell radius, 
for the ``top'' solution of set II; for all the other sets of solutions the 
behavior is very similar. The pion decay constant decreases by increasing
the density, going in the direction of a partial restoration of chiral
symmetry, although the lack of solutions beyond roughly the standard
saturation density of nuclear matter, prevents one, --- at the present stage of
development of the model, --- from drawing firmer conclusions. 
A reduction of $f_\pi$ in matter is found both in linear sigma and 
Nambu-Jona-Lasinio models (see, e.~g., Ref.~\cite{Bir94} for a list of
references). 

In Fig.~\ref{fig:sqrm} the isoscalar mean radius is plotted as a function of 
the cell radius, for the ``top'' (solid) and ``bottom'' (dashed) solutions of 
set II; the other sets of solutions give very similar results. 
We observe that this quantity is extremely sensitive to the choice of ``top''
or ``bottom'' boundary conditions, the reason being the same as already noticed
when discussing Fig.~\ref{fig:rhotb}: The ``bottom'' solution in fact 
corresponds to a configuration in which the quarks are more loosely packed 
inside the cell.
One could in principle define a mean radius averaged over the solutions 
distributed in the band, assuming, for instance, a uniform distribution. 
From Fig.~\ref{fig:sqrm} one would expect such an averaged radius to have a
moderate density dependence. However, as already mentioned, we expect
confinement to affect the band width, and expecially the ``bottom'' border
(actually, the upper end), thus unbalancing the distribution in favor of the
lower curve of Fig.~\ref{fig:sqrm} and yielding a shrinking of the
soliton. This would be at variance with findings of other models.

For completeness we display in Figs.~\ref{fig:theta123}, \ref{fig:uv123} and 
\ref{fig:rho123}, as a function of the distance from the center of the WS cell
and for three values of $R$, the chiral angle, the large and the small
components of the Dirac spinor and the baryon density, respectively, using the
``top'' solutions of sets II and III. Similar results apply for the other
solutions. 

A quantity which is more sensitive to the choice of boundary conditions is 
$g_A$, as one can see in Fig.~\ref{fig:ga}. The local mesonic component varies
noticeably in the different cases, even at large $R$. 
In Table \ref{tab:gA} we display the value of the different components of $g_A$
at $R=5$~fm, compared to the results for the free soliton case\footnote{ As
noted by the authors of Ref.~\cite{Adj92}, the two-point approximation to the
effective action works for $g_A$ to a lesser extent than for the energy and the
apparent agreement with the experimental value should be regarded as 
accidental, since the self-consistent calculation (see, e.~g.,
Ref.~\cite{Wak91}) gives a smaller value.}.
One clearly sees that the quark and non-local mesonic components have already
reached the asymptotic value, whereas the local mesonic term is still higher.
The reason for this behavior can be traced back to the slow decay of the chiral
angle in the chiral limit ($\theta(r) \approx 1/r^2$), which allows sizeable 
contributions from large distances. Although in principle one could continue
the WS calculation at larger radii, one would then need to increase the number
of states included in the orthonormal basis employed in the expansion.
However, it is likely that, when chiral symmetry is explicitly broken, the 
exponential decay of the chiral angle would grant a faster convergence and
suppress the sensitivity to the boundary conditions.
Increasing the density, the mesonic contributions to $g_A$, both local and
non-local, rapidly decrease and most of the strength is now carried by the 
quarks alone. 
\begin{table}
\caption{ }
\begin{tabular}{ccccc}
  & quark & local mesonic & non-local mesonic & total \\
\tableline
free & 1.07 & 0.50 & -0.35 & 1.22\\
II   & 1.06 & 0.76  & -0.34  & 1.48 \\
III  & 1.06 & 1.03  & -0.34 &  1.76 \\
\end{tabular}
\label{tab:gA}
\caption{ The axial coupling constant $g_A$ in free space and in the
Wigner-Seitz approximation at $R=5$~fm, for the ``top'' boundary conditions of 
sets II and III.}
\end{table}

\section{ Outlook and perspectives }
\label{sec:concl}

In this paper we have applied the Wigner-Seitz approximation to the 
chiral quark-soliton model of the nucleon. 
This model, complemented by the WS approximation, provides a simple yet 
interesting framework for studying possible modifications of the nucleon 
properties in nuclear matter. 
It is important to remark that we are actually dealing with a parameter-free
model, since the only free parameters (the constituent quark mass $M$ and the
regularization scale) have been fixed in free space.
Note that $M$ could, in principle, be calculated dynamically, as done, e.~g.,
in the Nambu-Jona-Lasinio model \cite{Rip97}; this would require the solution 
of a gap equation and would probably lead to an in-medium suppression of $M$, 
similar to what is found for $f_\pi$\footnote{By increasing the density, only
virtual states of higher energy can be excited and, eventually, in a very dense
system, only states lying well above the typical momentum cutoff would be 
available: The quark condensate, $\langle \overline{q} q \rangle$, is thus 
expected to decrease at finite densities.}.

The same method of accounting for in-medium effects has already been applied to
a number of microscopic models of the nucleon
\cite{Wus87,Amo98,Rei85,Bir88,Web98,Ban85,Gle86,Hah87,Joh96,Joh97,Joh98}.
With respect to previous analyses, we have for the first time consistently
calculated effects stemming from the Dirac sea, i.~e., from excitations of
virtual quark--antiquark pairs, including their dependence upon the density of
the system.

Vacuum fluctuations manifest themselves in two ways. Firstly, by giving rise to
a pion decay constant dependent on the distance from the center of the bag,
which in turn modifies the pion kinetic contribution with respect to chiral
models where the pion is explicitly included at the classical level
\cite{Ban85,Gle86,Hah87,Web98} (see Eqs.~(\ref{eq:EkinWS}) and
(\ref{eq:fpiWS})): The contribution to the energy from this term is however
qualitatively similar to the one of previous analyses
\cite{Ban85,Gle86,Hah87,Web98}. Secondly, vacuum fluctuations generate a new
non-local, {\em attractive} contribution, whose density dependence, albeit
moderate, is however relevant in order to bind the system, given the
compensation one observes with increasing density between the valence quark and
kinetic pion contributions.

Another effect that has usually been neglected in previous calculations is the
spurious center-of-mass motion: We have found it to be important not only to
give, of course, a more realistic estimate of the energy, but also to determine
the relative position of the top and bottom ends of the quark energy band 
in the medium. These two levels correspond to specific boundary conditions and
we have found the role of the boundary conditions generating the upper and
lower levels to be exchanged with respect to previous calculations.
Note that the intersection of an occupied band with an empty one is often
interpreted as the onset of color superconductivity.

We have also explored the effects stemming from different choices of boundary
conditions. All the cases discussed above display a similar qualitative
behavior, although there are definitely quantitative differences, especially
for the axial coupling constant, which is however very much affected by the
long-range behavior of the chiral field in the (massless pion) chiral limit.
Also the two regularization schemes we have adopted turned out to give
qualitatively similar results.

We have calculated a few physical quantities, such as $f_\pi$ and $\langle
r^2\rangle_{I=0}$. The pion decay constant has been found to decrease with
increasing density, pointing to a partial restoration of chiral symmetry in the
medium. The isoscalar mean square radius, on the other hand, has been found to
depend heavily on the position in the energy band, leaving open the question
whether an average nucleon will swell or shrink, given the present
uncertainties in assessing the band structure.

Unlike previous calculations with chiral models \cite{Ban85,Gle86,Hah87,Web98},
we have been able to find binding, but at a density much lower than the
standard saturation density of nuclear matter; moreover, solutions of the 
equations of motion disappear roughly below the latter density.
The authors of Ref.~\cite{Hah87} have been able to find solutions at high
densities, --- still keeping realistic values for the constituent quark mass
parameter $M$, --- by incorporating in their chiral Lagrangian terms of higher
order in the pion field. In the present model this would correspond to dropping
the two-point approximation to the effective action in evaluating vacuum
fluctuations. This is probably the most needed development of the present
calculation, --- employing, e.~g., the numerical algorithm of
Ref.~\cite{Kah84}, --- since only in the free case it has been explicitly 
verified that the two-point approximation works reasonably well \cite{Adj92}.

Finally, another feature that is lacking at the present stage is the projection
of good spin-isospin quantum numbers out of the hedgehog state. Since the 
momentum of inertia is, in general, density-dependent, the projection is likely
to affect the position of the minimum in the energy.

\acknowledgements

We would like to thank Prof. J. D. Walecka for many useful discussions.
This work has been supported in part by DOE grant DE-FG05-94ER40829.

\appendix

\section{The propagator $K(q)$}
\label{app:Kq}

We derive here the expression for the propagator $K(q)$, introduced in
Sect.~\ref{subsec:effact2}.
As a first step we perform the angular integrals in Eq.~(\ref{eq:Kqr}),
getting
\begin{eqnarray}
  K(q) &=& \frac{1}{2\pi^4} \int_0^\infty dk\,k^2 \int_0^\infty dk'\,{k'}^2
    \frac{\pi}{E_k E_{k'}} \frac{C\left[\left(k+k'\right)/2 
    \right]}{E_k+E_{k'}}  \int_0^\infty dr\,r^2 j_0(qr) j_0(kr) j_0(k'r) . 
\end{eqnarray}
The inner integral can be done analytically \cite{Grad80}, yielding
\begin{eqnarray}
  \int_0^\infty dr\,r^2 j_0(qr) j_0(kr) j_0(k'r) &=& 
    \left(\frac{\pi}{2}\right)^{3/2}
    \frac{1}{\sqrt{q k k'}} \int_0^\infty dr\,\sqrt{r} J_{1/2}(qr) J_{1/2}(kr)
    J_{1/2}(k'r) \nonumber \\
  &=& \frac{\pi}{4} \frac{\Delta(q,k,k')}{q k k'} ,
\end{eqnarray}
where the function $\Delta(x,y,z)$ vanishes whenever it is not possible to
build a triangle of sides $x$, $y$ and $z$, and is equal to $1$ otherwise.

One then obtains
\begin{equation}
  K(q) = \frac{1}{8 \pi^2 q} \int_0^\infty dk\,\frac{k}{E_k} 
           \int_{|k-q|}^{k+q} dk'\,\frac{k'}{E_{k'}}
           \frac{C\left[\left(k+k'\right)/2 \right]}{E_k+E_{k'}} .
\end{equation}

\section{ A basis for flat functions}
\label{app:basis}

In free space ($R \to \infty$) an orthonormal and complete set of states can be
chosen as
\begin{equation}
  \Psi_{k_0klm} (x) = \frac{1}{\pi} e^{i k_0 x_0} Y_l^m(\Omega) j_l(k r) .
\label{eq:freebasis}
\end{equation}
We want to build an orthonormal and complete basis inside a sphere
of radius $R$, having the same form of (\ref{eq:freebasis}). 
While in that case the momentum could take a continuum set of values, now
only a discrete set of momenta will be allowed.

We start considering the equations for the spherical Bessel functions,
corresponding to two different momenta, $k_1 = \alpha/R$ and
$k_2 = \beta/R$: 
\begin{eqnarray}
  \frac{d}{dr} \left[ r^2 \frac{d}{d r} j_l \left(\frac{\alpha r}{R}\right) 
    \right] + \left[ \left(\frac{\alpha r}{R}\right)^2 - l (l+1) \right] 
    j_l\left(\frac{\alpha r}{R}\right) &=& 0 \nonumber \\
  \frac{d}{dr} \left[ r^2 \frac{d}{d r} j_l\left(\frac{\beta r}{R}\right) 
    \right] + \left[ \left(\frac{\beta r}{R}\right)^2 - l (l+1) \right] 
    j_l\left(\frac{\beta r}{R}\right) &=& 0  . 
\end{eqnarray}
Multiplying both expressions by $j_l (\alpha r/R)$ and $j_l (\beta r/R)$, 
respectively, integrating over $r$ between $0$ and $R$, taking the difference 
of the two equations and, finally, integrating by parts, one gets
\begin{equation}
  - \int_0^R \left(\alpha^2 - \beta^2 \right) \frac{r^2}{R^2} 
    j_l\left(\frac{\alpha r}{R}\right) j_l\left(\frac{\beta r}{R}\right) = 
    R \left[ \alpha j'_l(\alpha) j_l(\beta)  
    - \beta j'_l(\beta) j_l(\alpha) \right]  
\label{eq:ortho}
\end{equation}
The above equation states the orthonormality of the elements of the basis, 
provided that the following condition is met:
\begin{equation}
  \alpha j'_l(\alpha) j_l(\beta) - \beta j'_l(\beta) j_l(\alpha) = 0,
    \quad \alpha \neq \beta . 
\end{equation}
The choice of the boundary conditions fulfilled by the elements of the
basis will therefore be constrained by this requirement. It is easy to
convince oneself that three different possibilities are available:
\begin{mathletters}
\begin{eqnarray}
  j_l(\alpha) &=& 0 \\
  j_l'(\alpha) &=& 0 \\
  \label{eq:bceta}
  \alpha j_l'(\alpha) &=& \eta j_l(\alpha) ,
\end{eqnarray}
\end{mathletters}
where $\eta$ will is a constant parameter. In the limit $\eta = 0$ and 
$\eta \to \infty$ the first two cases are recovered, respectively.

Notice that, in the limit $R \to \infty$, --- where the modes form a continuum,
--- all these boundary conditions become equivalent.

We then introduce the normalized functions
\begin{eqnarray}
  \rho_{\alpha_l}(r) &\equiv& 
    \kappa_{\alpha_l} j_l\left(\frac{\alpha_l r}{R}\right) \nonumber \\
  \kappa_{\alpha_l}^2 &=& \frac{2}{R^3} \left[ \alpha_l^2 j_l'(\alpha_l)^2 -
  j_l(\alpha_l) \left( -2 \alpha_l j_l'(\alpha_l) + ( l (l+1) - \alpha_l^2) 
  j_l(\alpha_l) \right) \right]^{-1}  ,
\end{eqnarray}
such that
\begin{equation}
  \int_0^R dr \ r^2  \rho_{\alpha_l}(r) \rho_{\beta_l}(r) =  
    \delta_{\alpha_l \beta_l} .
\end{equation}
The completeness of the basis allows one to write the following representation 
for the Dirac delta function inside the WS cell:
\begin{equation}
  \sum_{\alpha_l} \rho_{\alpha_l} (r) \rho_{\alpha_l} (r') = 
    \frac{\delta (r-r')}{r^2} .
\end{equation}
It is simple to derive an expression for the Green's function of the operator 
$(\Box+M^2)$ in the WS cell. It reads
\begin{equation}
  G(x,x') = \sum_{l,m} \int_{-\infty}^\infty \frac{dk_0}{2 \pi} \sum_{\alpha_l}
    Y_l^m(\Omega) Y_l^{m *} (\Omega') e^{-i k_0 (x_0 - x'_0)}  
    \frac{\rho_{\alpha_l}(r) \rho_{\alpha_l}(r')}{\alpha_l^2/R^2 + M^2 - k_0^2}
    \nonumber .
\end{equation}
The flat basis is obtained, as we said above, in the limit $\eta\to0$; 
in this case, however, the lowest energy mode, --- corresponding to zero 
momentum for $l=0$, can only be obtained for a strictly vanishing $\eta$ (in
fact, it corresponds to a negative root of Eq.~(\ref{eq:bceta}) at finite 
values of $\eta$). This means that the strength of this mode has to be 
redistributed, for finite $\eta$, over all the other modes. 
Let us call $\beta$ and $\tilde{\beta}$ the modes such that
\begin{equation}
  j'_l(\beta)=0 , \quad 
    \tilde{\beta} j_l'(\tilde{\beta}) = \eta j_l(\tilde{\beta}) .
\end{equation}
There is a one-to-one correspondence between the modes in the two bases, in 
such a way that for each $\beta\ne0$ one can define 
$\tilde{\beta}=\beta-\epsilon_\beta$, where $\epsilon_\beta\ll 1$ if 
$\eta\ll 1$.

Writing down the transformation from one basis to the other, namely
\begin{equation}
  \rho_\beta(r) = \sum_{\tilde{\beta}} c_{\beta \tilde{\beta}} 
    \tilde{\rho}_{\tilde{\beta}}(r) ,
\label{eq:four-bes}
\end{equation}
with some algebra one can verify that, for $l=0$ and $\beta=0$, one has 
\begin{equation}
  c_{0\tilde{\beta}} = -\frac{\sqrt{6}}{\tilde{\beta}^2} \epsilon_\beta ,
\end{equation}
which indeed proves that the zero mode is now redistributed over all the other
modes, in the quasi-flat basis.

Calculations in the two bases are equivalent (to order $\eta$) only when 
considering  convergent quantities. Quantities that need regularization 
are therefore different in the two bases, since only a finite number of modes 
will contribute. The flat basis is the one we have chosen in the calculations 
of this paper, since it is the most convenient in order to impose the 
physically motivated boundary conditions discussed in the text.
The zero mode, however, gives rise to divergences when $R\to0$ and
one could in principle eliminate this problem by employing a quasi-flat basis 
infinitesimally close to the flat one. On the other hand, the very fact that 
for regularized quantities only a finite number of modes is relevant, allows us
to keep the flat basis and simply discard the zero mode, since the error 
introduced in this way will be infinitesimally small.

Finally, we display in Table~\ref{tab:WSrules} a few simple rules that allow 
one to rewrite in the WS basis quantities expressed in the free basis 
(\ref{eq:freebasis}).
\begin{table}
\begin{center}
\begin{tabular}{ccc}
  free space&~&Wigner-Seitz boundary conditions \\ 
  \tableline
  $\int_0^\infty dk \ k^2$  & $\longrightarrow$ & $(\pi/2)\sum_{\alpha_l}$ \\  
  $k$                       & $\longrightarrow$ & $\alpha_l/R$ \\
  $j_l (k r)$               & $\longrightarrow$ & $\rho_{\alpha_l}(r)$ 
\end{tabular}
\caption{ Rules for going from free space to the Wigner-Seitz cell. }
\label{tab:WSrules}
\end{center}
\end{table}

\end{document}